\documentclass[%
reprint,
 amsmath,amssymb,
 aps,
]{revtex4-2}

\usepackage{amsmath}
\usepackage{graphicx}
\usepackage{dcolumn}
\usepackage{bm}
\usepackage{amsmath}
\usepackage{float}
\usepackage{mathtools}  
\usepackage{xfrac} 

\usepackage{lineno} 

\switchlinenumbers

 \usepackage{hyperref}
 \hypersetup{
     colorlinks=true,
     linkcolor=blue,
     filecolor=blue,
     citecolor = blue,      
     urlcolor=blue,
     }
     \usepackage{tikz,xcolor,hyperref}

\definecolor{lime}{HTML}{A6CE39}
\DeclareRobustCommand{\orcidicon}{%
	\begin{tikzpicture}
	\draw[lime, fill=lime] (0,0) 
	circle [radius=0.16] 
	node[white] {{\fontfamily{qag}\selectfont \tiny ID}};
	\draw[white, fill=white] (-0.0625,0.095) 
	circle [radius=0.007];
	\end{tikzpicture}
	\hspace{-3.5mm}
}

\foreach \x in {A, ..., Z}{%
	\expandafter\xdef\csname orcid\x\endcsname{\noexpand\href{https://orcid.org/\csname orcidauthor\x\endcsname}{\noexpand\orcidicon}}
}

\begin{document}
\preprint{APS/123-QED}



\title{Supercavity  Modes in Stacked Identical Mie-resonant Metasurfaces }

\author{Xia Zhang\orcidA{}}
\email{zhangxia1@mail.neu.edu.cn}
\author{Xin Zhang\orcidC{}}
\email{zhangxin@mail.neu.edu.cn}
 \affiliation{College of Sciences, Northeastern University, Shenyang 110819, China}
\author{A. Louise Bradley\orcidB{}}%
 \email{bradlel@tcd.ie}
\affiliation{%
School of Physics, CRANN and AMBER, Trinity College Dublin, Dublin, Ireland and IPIC, Tyndall National Institute, T12 R5CP Cork, Ireland.
}%

\date{\today}

\begin{abstract}

Modes with a high-$Q$ factor are crucial for photonic metadevices with advanced functionalities. In sharp contrast to recent techniques which generate a supercavity mode by bound states in the continuum via symmetry breaking, we reveal a general and new route, by stacking two parallel and identical Mie-resonant metasurfaces with an air separation. The supercavity mode can be designed by the established theoretical model by overlapping Mie resonant modes with tailor-made Fabry-P\'{e}rot modes. 
The simplified system, with free-space field concentration which can exist in plane as well as out of plane of the metasurfaces, provides for ease of integration with added matter, creating exciting different opportunities for the study of fundamental light-matter coupling.
This work deepens our understanding of light manipulation using metasurfaces. It paves a different and general route for generating supercavity modes which can be easily engineered and controlled for different applications of all dielectric metaoptics.

\end{abstract}

\maketitle
\section{Introduction}

High-$Q$ cavities with field concentration in localized modes is crucial for a plethora of applications in optics and photonics, such as lasing \cite{Bhattacharya2014,de2009stimulated}, sensing \cite{doi:10.1063/1.4978672}, nonlinear harmonic generation \cite{michaeli2017nonlinear,koshelev2020subwavelength}, and Raman scattering \cite{PhysRevLett.126.123201}. Plasmonic resonators have been the platform of choice for field concentration in subwavelength volumes \cite{schuller2010plasmonics,hu2018experimental}. However, high-$Q$ resonances require low loss, which is one inevitable obstacle for  plasmonic structures.  Low-loss dielectric structures with high refractive index materials have become a topic of intensive research as a new route for high-$Q$ resonant modes by employing Mie-type resonances \cite{Won2019,Kivshar2022}.
Such Mie-type resonant modes
are dominated by the radiation damping. Furthermore, the involvement of higher-order coherent electric and magnetic modes enables flexible manipulation of the incident electromagnetic wave. Metasurfaces have been applied in a range of meta-devices, such as magnetic mirrors, lenses, waveplates and holograms \cite{Yu2014,PhysRevLett.119.123902, zhang2021wide}.

Mie resonances generally do not provide a high Q-factor. However, there are several methods to achieve high $Q$ factors employing dielectric nanostructures, such as by confining waves by introducing a defect in photonic crystals \cite{quan2010photonic, akahane2003high}, exploiting whispering gallery modes \cite{PhysRevLett.85.74,kuo2014second}, employing collective lattice modes \cite{Zhao:21} and more recently via quasi bound states in the continuum (quasi-BIC) using an individual nanoresonator or in an array-type metasurface\cite{PhysRevLett.119.243901,PhysRevLett.121.193903}. The symmetry-protected quasi-BIC eigenmodes only become accessible when a symmetry mismatch exists. This is typically realized via an oblique angle of incidence or geometric symmetry breaking \cite{doeleman2018experimental,PhysRevLett.121.193903}.

Dielectric metasurfaces have attracted much attention due to their unprecedented capabilities for controlling the amplitude, phase, and polarization \cite{PhysRevApplied.9.034005,yu2015high} of the scattered electromagnetic radiation. The dielectric metasurface can be treated as an inhomogeneous sheet of interfering Mie-type electric and magnetic resonances \cite{evlyukhin2016optical, PhysRevApplied.9.034005, shamkhi2019transparency}. While the reflection and refraction of a planar dielectric layer can be described by the Fresnel equations and Snell's law \cite{Chen2016}, a single dielectric metasurface can be used to arbitrarily tailor the reflectance, $R$ and transmittance, $T$  \cite{Lee2021, arbabi2015dielectric,zhang2021wide}. Furthermore, the spectral response of an isolated nanoparticle, as well as the metasurface, can be well described by multipolar mode decomposition in Cartesian coordinates \cite{shamkhi2019transparency,evlyukhin2016optical,terekhov2017multipolar,zhang2021wide}.  The well-defined spectral responses enable a deep understanding of the underpinning physics and therefore offer a route to explore more exotic optical and photonic designs \cite{Yu2014, PhysRevLett.119.123902}. 

 A Fabry-P\'{e}rot mode is formed when two mirrors are placed together, with examples including metallic mirrors, distributed Bragg reflectors, parallel-aligned subwavelength photonic crystals or dielectric gratings \cite{suh2003displacement,marinica2008bound,wu2022tailoring}. Mirrors can also be generated by dielectric resonators  \cite{Liu2014mirror,PhysRevLett.119.123902} and by extension, a Fabry-P\'{e}rot mode can also be created for light confinement by using two dielectric metasurface-based mirrors.  Herein we report on a supercavity formed by two dielectric metasurfaces in a stacked geometry. It is demonstrated that a very high Q-factor $\sim10^5$ and field enhancement can be achieved due to the coherent interplay of the Mie resonances of the dielectric metasurfaces forming a Fabry-P\'{e}rot bound states in the continuum mode.

 The paper is structured as follows. In Sec.~\ref{sec2}, we first review the multipolar modes driving the resonant monolayer dielectric metasurface. Then a semi-analytical theoretical model is developed based on the transfer matrix method assuming that the monolayer metasurface is treated as a sheet of coherent multipolar modes while the stacked metasurfaces are decoupled or weakly-coupled. In Sec.~\ref{sec3}, we investigate the resonant modes of monolayer and stacking metasurfaces in the near- and far-field, validate the proposed theoretical model in the far-field, and quantitatively interpret the numerically simulated resonant energies and lineshapes using the temporal coupled-mode theory. The origin of the observed high-$Q$ resonant modes from the simulations is further revealed. In Sec.~\ref{sec4}, conclusions are presented.

\section{Theoretical Model}\label{sec2}

\begin{figure}[htbp]
\includegraphics[width=1\linewidth]{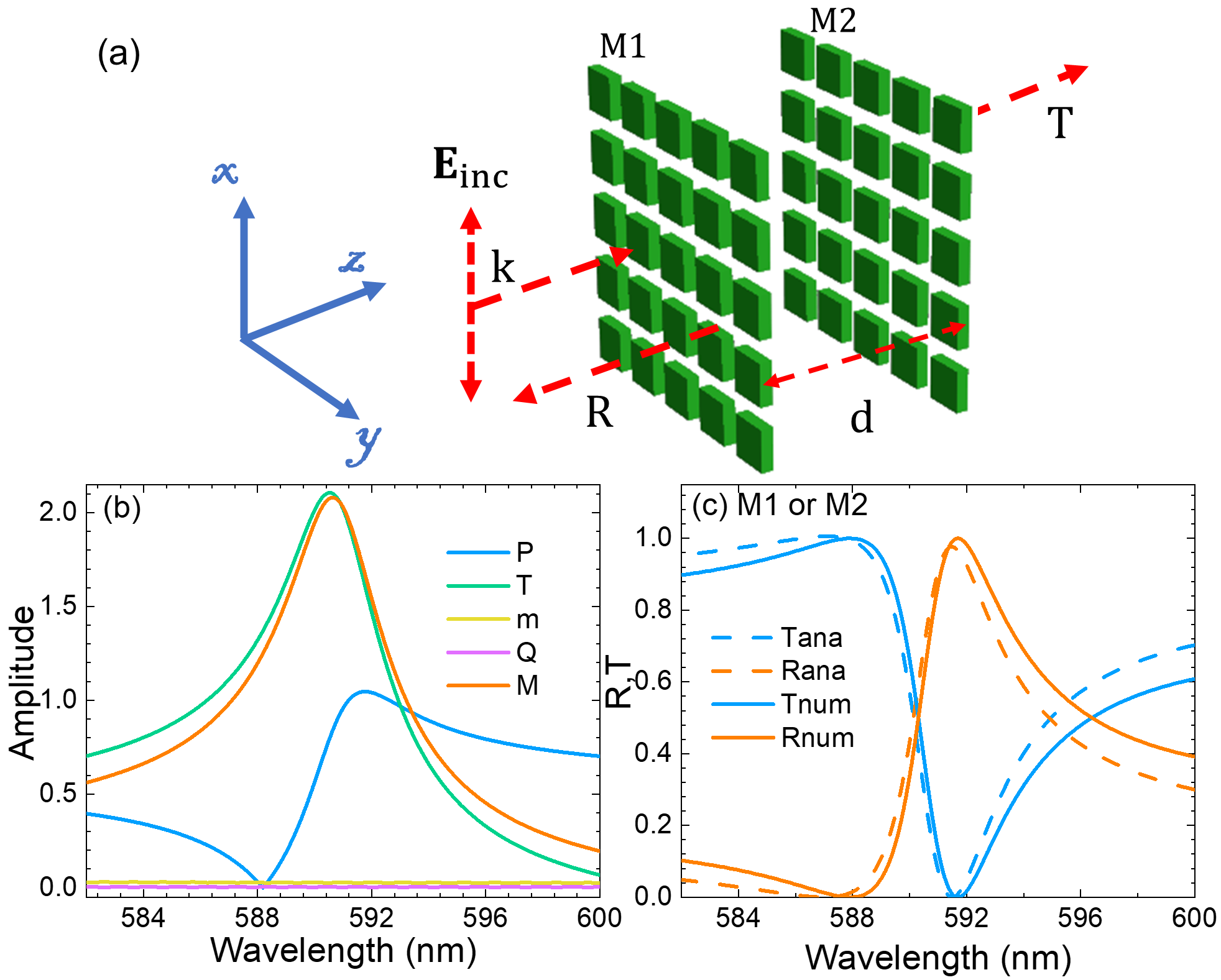}
\caption{\label{fig:sch}(a) Schematic of the geometry containing two parallel identical metasurfaces (M1 and M2) in free-space, normally illuminated by a plane wave, $\bm{E}_{inc}=E_0e^{i(k_0 z-\omega t)}\bm{x}$. The metasurfaces are perfectly aligned 2D periodic arrays of nanocubes with period $\Lambda_x$ and $\Lambda_y$ along $x$ and $y$ directions, and thickness of $t$. The edge length of the cube is $l_x=l_y =430$ nm, and $t=60$ nm, $\Lambda_x$ = $\Lambda_y$ = 500 nm. The two metasurfaces have center-to-center distance, $d$ with the air-gap of thickness, s, where $d=s+t$. (b) Amplitude of the multipole moments contributing to reflectance and transmittance calculated according to Eqs. (\ref{eq:rt}) and (\ref{eq:MD}), including electric dipole, $r_P$,  toroidal dipole, $r_T$, magnetic dipole $r_m$, electric quadrupole, $r_Q$ and magnetic quadrupole $r_M$. (c) Numerically simulated (num) and semi-analytically calculated (ana) reflectance (R) and transmittance (T) coefficients.  }
\end{figure}

Two isolated, identical metasurfaces, labeled as M1 and M2, respectively, are depicted in Fig.~\ref{fig:sch}. The metasurfaces are comprised of a square array of nanocubes.  Within this work it is assumed that the nanocube material is titanium dioxide (TiO$_2$)), with the refractive index taken from 
the experimental data in Ref. \cite{Sarkar2019}. TiO$_2$ is lossless in the visible wavelength range. The real and imaginary values of the refractive index data of TiO$_2$ are given in Fig.~\ref{n} in Appendix.
The lattice constants in the $x$ and $y$ directions are $\Lambda_x$ and $\Lambda_y$ with $\Lambda_x$ = $\Lambda_y$ = 500 nm. M1 and M2 are separated by air with the distance of $d$ as illustrated. Each cube has the edge length, $l_x=l_y =430$ nm, thickness $t=60$ nm. The center-to-center distance between two metasurfaces is $d$ with $d=s+t$, where $s$ denotes the air separation. For one isolated metasurface, the transmission coefficient and reflection coefficient are expressed as $t$ and $r$ following Ref. \cite{shamkhi2019transparency,zhang2020constructive,zhang2021wide} up to quadrupolar terms as

\begin{equation}{\label{eq:rt}}
\begin{split}
&{r}=\frac{ik_0}{2E_0A\epsilon_0}(\textbf{P}_x +ik\textbf{T}_x-\frac{1}{c}\textbf{m}_y+\frac{ik_0}{6}\textbf{Q}_{xz}-\frac{ik_0}{2c}\textbf{M}_{yz}),
\\&{t}=1+\frac{ik_0}{2E_0A\epsilon_0}(\textbf{P}_x +ik\textbf{T}_x+\frac{1}{c}\textbf{m}_y-\frac{ik_0}{6}\textbf{Q}_{xz}-\frac{ik_0}{2c}\textbf{M}_{yz}),
\end{split}
\end{equation}
where $A$ is the area of a unit cell of the metasurface.  $\mathbf{P}_x$, $\mathbf{T}_x$  $\mathbf{m}_y$, $\mathbf{Q}_{xz}$ and $\mathbf{M}_{yz}$ refer to the effective electric dipole (ED),  toroidal dipole (TD), magnetic dipole (MD), electric quadrupole (EQ) and magnetic quadrupole (MQ), respectively under $x$-polarized incident light. The corresponding amplitude of each mode is $r_P$, $r_T$, $r_m$, $r_Q$ and $r_M$, respectively. These multipolar mode contributions are calculated by integrating the electric field in one unit cell following Refs. \cite{evlyukhin2016optical,terekhov2017multipolar,zhang2021wide}.

\begin{equation}{\label{eq:MD}}
\begin{split}
&\textbf{P}=\int \epsilon_0(\epsilon_{m}-1)\textbf{E}(\textbf{r})d\textbf{r},
\\& { \textbf{T}=\frac{-i\omega}{10c}\int\epsilon_0(\epsilon_{m}-1)[(\textbf{r}\cdot\textbf{E}(\textbf{r}))\textbf{r}-2\textbf{r}^2\textbf{E}(\textbf{r})]d\textbf{r}},
\\&\textbf{m}=-\frac{i\omega}{2}\int\epsilon_0(\epsilon_{m}-1)[\textbf{r}\times\textbf{E}(\textbf{r})]d\textbf{r},
\\&\textbf{Q}=3\int\epsilon_0(\epsilon_{m}-1) [\textbf{r}\textbf{E}(\textbf{r})+\textbf{E}(\textbf{r})\textbf{r}-\frac{2}{3}(\textbf{r}\cdot\textbf{E}(\textbf{r}))\hat{U}]d\textbf{r},
\\&\textbf{M}=\frac{\omega}{3i}\int\epsilon_0(\epsilon_{m}-1)[(\textbf{r}\times\textbf{E}(\textbf{r}))\textbf{r}+\textbf{r}(\textbf{r}\times\textbf{E}(\textbf{r}))]d\textbf{r},
\end{split}
\end{equation}
where $\textbf{r}$ is the coordinate vector with its origin placed at the center of the nanocube. $\textbf{E}(\textbf{r})$ is the total electric field inside the nanocube at different position. $\epsilon_0$ is the vacuum permittivity; $\epsilon_{m}$ is the relative dielectric permittivity of the metasurface. $c$ is the light speed in vacuum; $\hat{U}$ is the 3$\times$3 unity tensor; \textbf{P}, \textbf{T}, \textbf{m}, \textbf{Q} and \textbf{M} are the moments of ED, TD, MD, EQ and MQ respectively for under arbitrary polarization of incident light.

Next we consider the spectral responses of the stacked identical metasurfaces (M1 and M2) with a center-to-center separation, $d$. Generally, two mechanisms underpin the coupling between two stacked metasurfaces, near-field coupling and far-field coupling, with the key distinction between both coupling mechanisms dependent on the separation of the metasurfaces. The near-field coupling is caused by the overlap of the resonant mode patterns of each metasurface. The far-field coupling is dominated by the radiative loss of the single metasurface and the relative phase shift of the stacked metasurfaces \cite{luo2022wavy}. According to the transfer matrix formalism, each layer can be treated as a decoupled in-line optical element \cite{katsidis2002general, chen2012interference, babicheva2017reflection, zhang2020constructive}, which implies a prerequisite of application only in the far-field regime with negligible near-field coupling between the individual metasurfaces.

The light response of a single metasurface is specified by its reflection coefficient,  $r$ and transmission coefficient, $t$. The relative phase is very important for the stacked metasurfaces as discussed in Ref. \cite{Feng2020}. The incident light wave on M1 is $\bm{E}_{inc}=E_0e^{i(k_0 z-\omega t)}\bm{x}$.  When the same light wave arrives on M2, an extra phase factor exists, $\bm{E}_{inc}^{'}=E_0e^{ik_0 (z+d)-i\omega t}\bm{x}$=$\bm{E}_{inc}e^{ik_0 d}$. However, the reflection and transmission coefficient do not include this additional phase change, and therefore it is added as the propagation matrix, $P_{air}$. 
The corresponding reflectance and transmittance coefficients of M2 become $re^{ik_0 d}$ and $te^{ik_0 d}$. The transfer matrix for the stacked metasurfaces, $\rm T= T_{M1}P_{air}T_{M2}$ becomes 

\begin{equation}
T=\frac{1}{t^2}\begin{bmatrix}
1 & -r\\
r & t^2-r^2
\end{bmatrix}
\begin{bmatrix}

e^{-ik_0d} & 0\\
0 & e^{ik_0d}
\end{bmatrix}
\begin{bmatrix}

1 & -r\\
r & t^2-r^2
\end{bmatrix}.
\end{equation}

The derived matrix elements are

\begin{equation}
\begin{split}
&T_{11}=\frac{1-r^2e^{i2k_0d}}{t^2e^{ik_0d}},
\\&T_{21}=\frac{r+r(t^2-r^2)e^{i2k_0d}}{t^2e^{ik_0d}}.
\end{split}
\end{equation}

The corresponding reflectance and transmittance are 
\begin{equation}{\label{eq:anar}}
R=|\frac{T_{21}}{T_{11}}|^2=|{\frac{r+r(t^2-r^2)e^{i2k_0d}}{1-r^2e^{2ik_0d}}|^2},
\end{equation}

\begin{equation}{\label{eq:anat}}
T={|\frac{1}{T_{11}}|}^2=|{\frac{t^2e^{ik_0d}}{1-r^2e^{i2k_0d}}|^2}.
\end{equation}

For a Fabry-P\'{e}rot cavity, while  $d=n\frac{\lambda}{2}$ is satisfied, the $n$th order Fabry-P\'{e}rot resonance exists with a standing wave formed in the cavity \cite{Rakhmanov2002}. According to Eq.~(\ref{eq:anar}) and Eq.~(\ref{eq:anat}), the spectral responses, R and T are periodic, which depends on $d$. While $2k_0d=2n\pi$, correspondingly, $d=n\frac{\lambda}{2}$ holds exactly, the same resonance appears. This confirms that the mode in origin has coherent contributions from a Fabry-P\'{e}rot mode and Mie resonant modes when the accumulating phase in a half round-trip is an integer multiple of 2$\pi$.

\section{Results and Discussions}\label{sec3}
\subsection{Resonant modes of a single metasurface}

To numerically calculate the R, T spectra of the metasurfaces, a finite-difference time-domain method using a commercial software (Ansys Lumerical FDTD) is applied. A mesh grid of 21.5 nm $\times$21.5 nm $\times$3 nm (corresponding to 20 electric field data points along $x$, $y$, and $z$ directions, respectively) is used and an auto-shutoff minimum 1e-9 is adopted considering the trade-off between numerical convergence, RAM capacity, and running time. Periodic conditions are applied on four sides of the unit cell for the metasurace with the incident linearly-polarized plane wave, $\bm{E}_{inc}=E_0e^{ik_0 z-i\omega t}\bm{x}$. The calculated three-dimensional electric field distribution within one unit cell, which takes account of the electric field of each resonator as well as the coupling between resonators, is used for multipolar decomposition.

First we consider the results for a single nanocube metasurace, M1 or M2. {The amplitudes of the decomposed multipolar modes according to Eq.~(\ref{eq:MD}) can be seen in Fig.~\ref{fig:sch} (b), which indicates that the dominant resonant modes of a single nanocube metasurface are ED, TD, and MQ, while MD and EQ are negligible. 
The numerical simulation and semi-analytical calculation for the reflectance and transmittance, R and T, respectively based on Eq.~(\ref{eq:rt}) are shown in Fig.~\ref{fig:sch} (c), and are found to be in good agreement.  The agreement proves that the coherent interplay of multipolar modes up to quadrupolar modes can reproduce the spectral properties of monolayer metasurface, which validates the dipole-quadrupole approximation for the multipole decomposition. }

\subsection{Resonant modes of two stacked metasurfaces:near-field and far-field coupling}

\begin{figure}[htbp]
\includegraphics[width=1\linewidth]{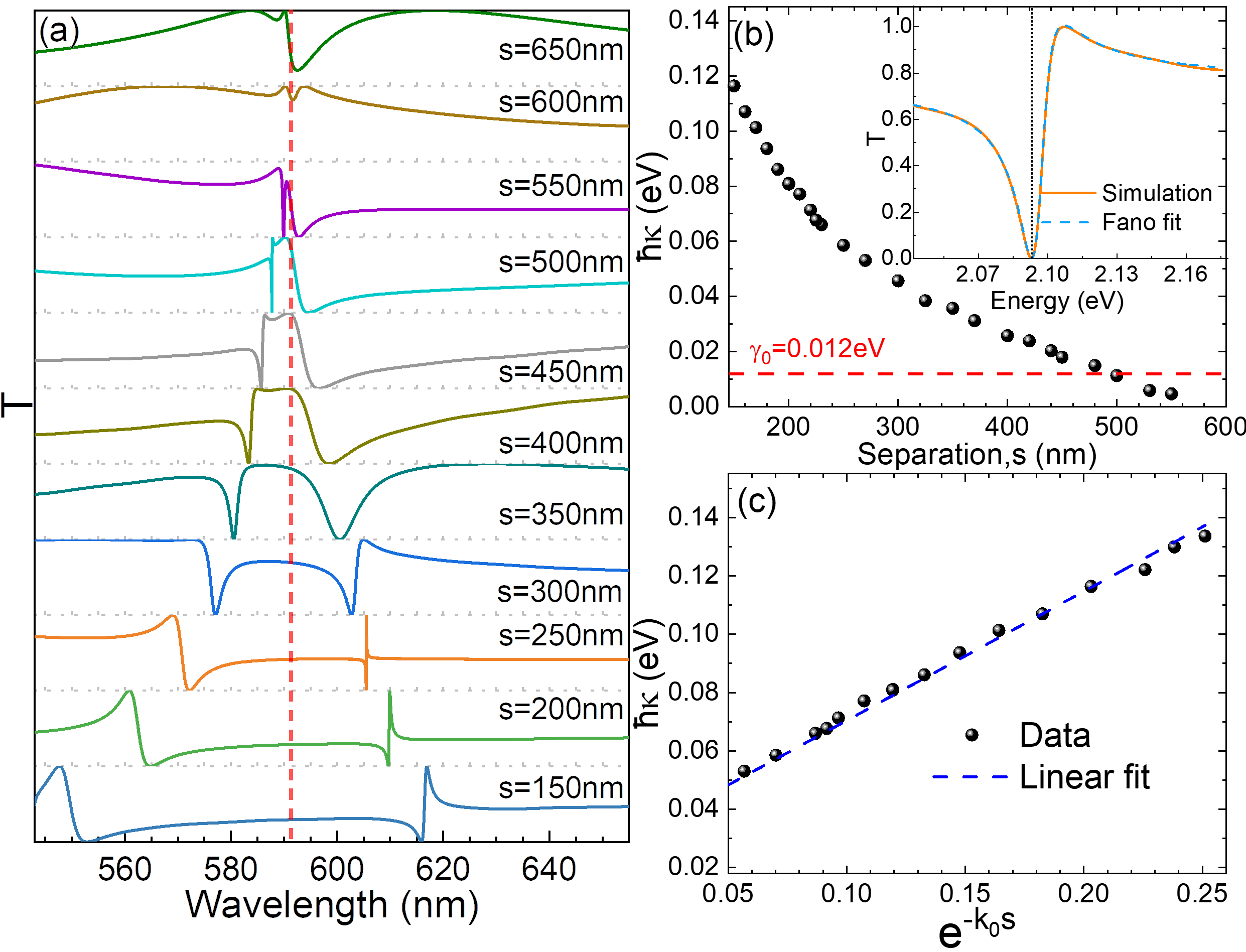}
\caption{\label{coupling} (a) Simulated transmittance spectra of the two stacked metasurfaces with the layer separation $s$ ranging from 150 nm to 650 nm. The red dashed line illustrates the resonance, which corresponds to the transmittance minimum at 591.6 nm, of the single monolayer metasurface. (b) The calculated coupling energy, $\hbar\kappa$ versus the metasurface separation $s$. Red dashed line illustrates the damping rate of the isolated single metasurface, which is $\hbar\gamma_0=0.012$ eV. (c) The near-field coupling energy, $\hbar\kappa$ versus $e^{-k_0s}$. The blue dashed line represents a linear fit to the data. }
\end{figure}

\begin{figure*}[htbp]
\includegraphics[width=0.95\linewidth]{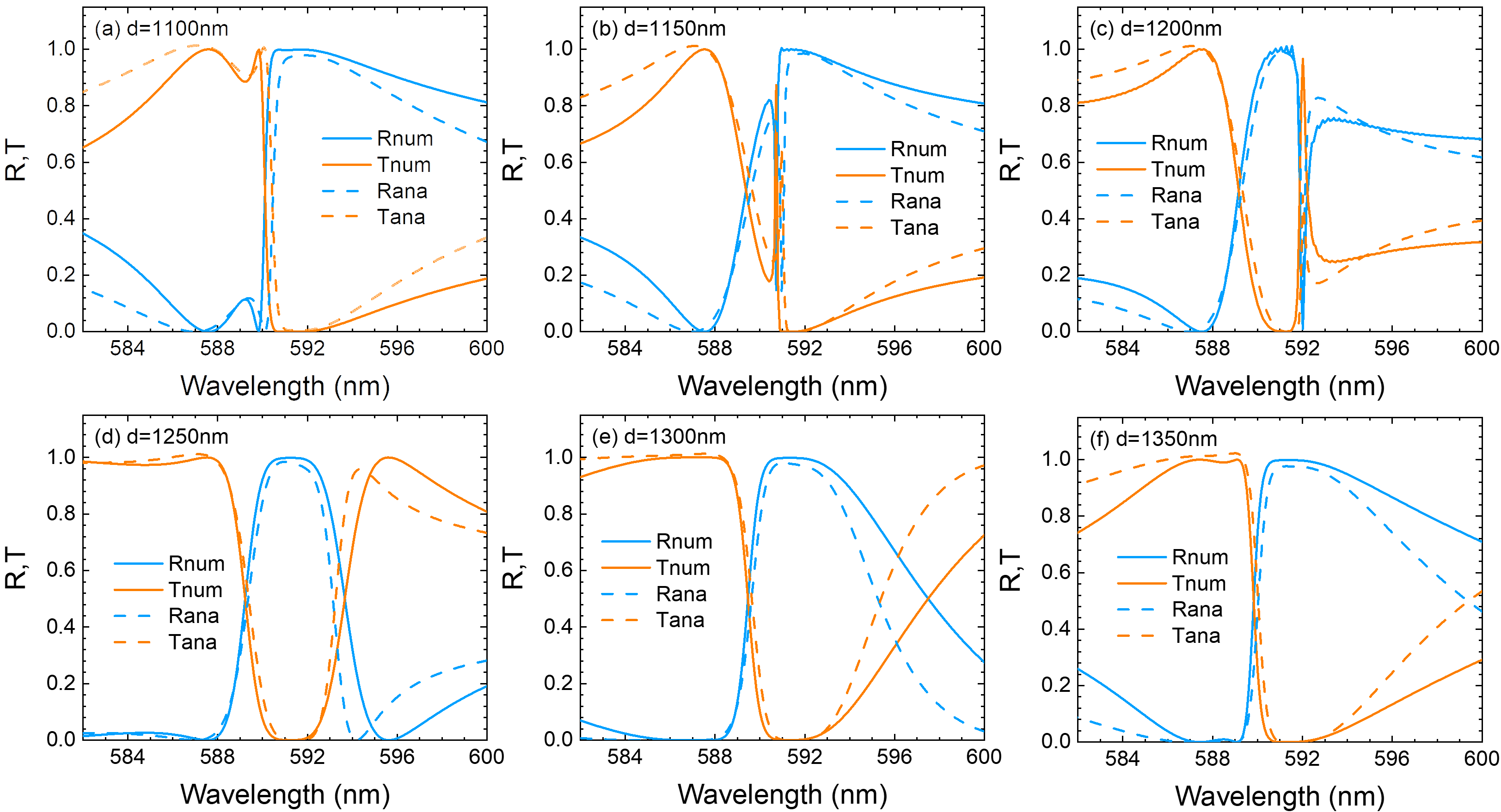}
\caption{\label{fig:fit}Numerically simulated and analytically calculated R, T curves at difference center-to-center separation, $d$. The edge length of the cube is $l_x=l_y =430$ nm, thickness $t=60$ nm. The stacked identical metasurfaces have the period, $\Lambda_x$ = $\Lambda_y$ = 500 nm.}
\end{figure*}

\begin{figure*}[htbp]
\includegraphics[width=0.9\linewidth]{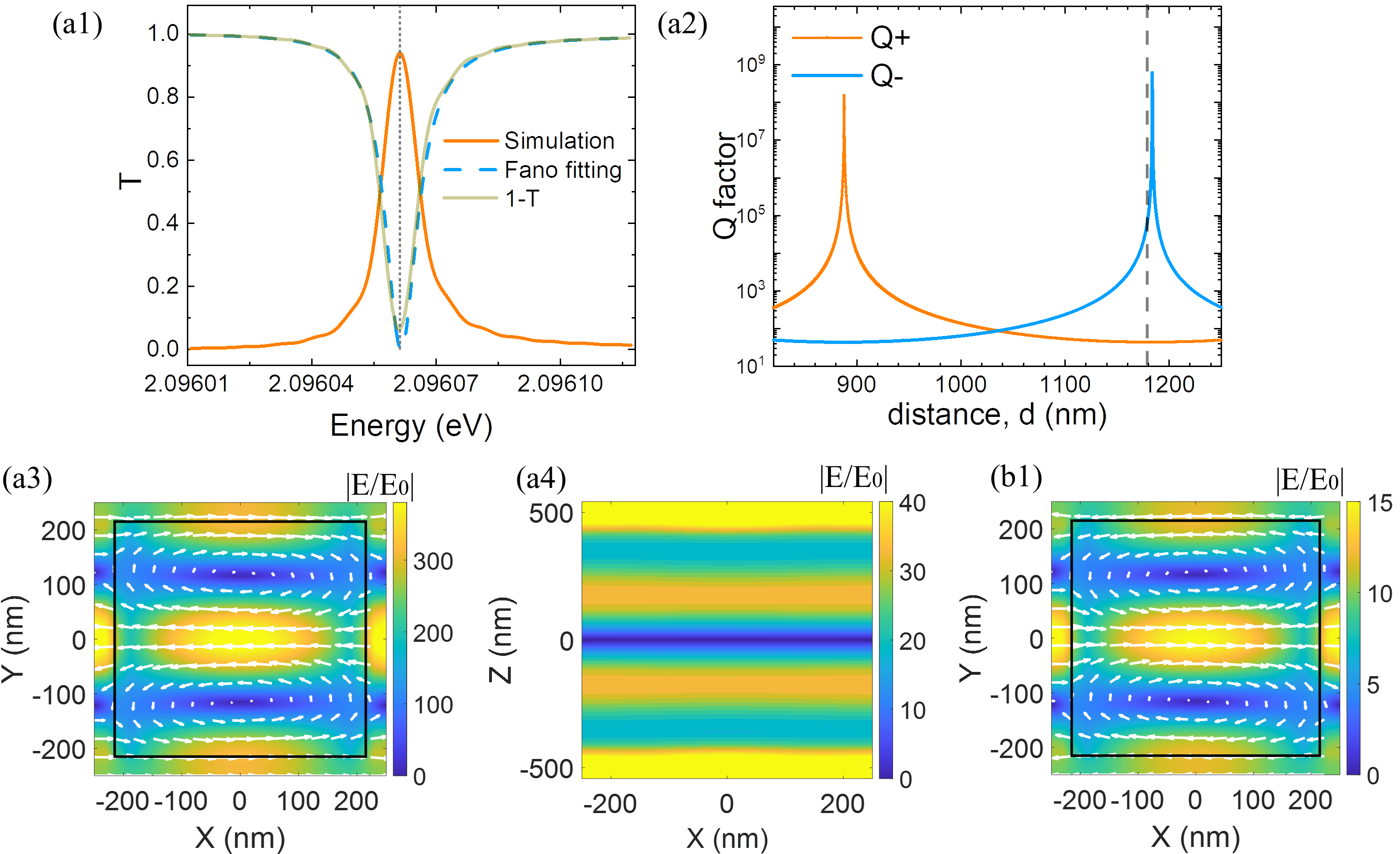}
\caption{\label{fig:highQ}Supercavity's spectral responses and field concentration ($d$ = 1180 nm). The transmittance, $T$, of the stacked identical metasurfaces (a1) with the Fano fit. The gray dashed line illustrates the resonance energy at approximately 2.09605 eV. The stacked identical metasurfaces, have the period, $\Lambda_x$ = $\Lambda_y$ = 500 nm, thickness, $t=60$ nm, The edge length of the cube is $l_x=l_y =430$ nm. (a2) Dependence of the calculated $Q$ factor of the resonant mode, $Q_{\pm}$, according to Eq.~\ref{eq:Q} on the distance between two metasurfaces, $d$. The gray dashed line illustrates the distance for $d$ = 1180 nm. (a3,b1)The amplitude ratio of the electric field, which is $x-y$ plane at the center of M1, with the corresponding field vectors at the inspected wavelength, 2.09605 eV ($\lambda$ = 591.6 nm) for M1 only (the spectra is shown in the inset of Fig.~\ref{coupling} (b)) and that in the stacked cases. (a4) The amplitude ratio of the electric field map ($x-z$ plane), with the $x-z$ field monitor spanning from edge-to-edge along $z$ direction and spans one unit cell along $x$ direction. }
\end{figure*}

Next we consider the spectral responses of the stacked identical metasurfaces (M1 and M2) with an air separation, $s$.  {As can be seen from the transmittance spectra in Fig.~\ref{coupling} (a) for the two stacked metasurfaces with varying air separation, a clear mode splitting with varying lineshapes is observed for $s$ over the range from 150 nm to 500 nm, in sharp contrast to the transmittance spectrum of a single metasurace, as seen in Fig.~\ref{fig:sch} (b). Near-field and far-field coupling underpin the spectral responses of the stacked two metasurfaces \cite{luo2022wavy}. To analytically interpret the simulation results, a general temporal coupled-mode theory is applied following Ref. \cite{fan2003temporal, wu2022tailoring}. The radiative damping rate of the resonant single metasurface is denoted as $\gamma$, the coupling rate or the coupling strength of two monolayer metasurfaces is denoted as $\kappa$. The Hamiltonian for the two coupled oscillators with the resonant energies, $E_1$ and $E_2$ and radiative damping rates, $\gamma_1$ and $\gamma_2$, respectively, can be written as}
\begin{equation}
H=
\begin{bmatrix}
E_1 & \hbar\kappa \\
\hbar\kappa & E_2 
\end{bmatrix} -i\hbar
\begin{bmatrix}
\gamma_1 & \sqrt{\gamma_1\gamma_2}e^{i\varphi} \\
\sqrt{\gamma_1\gamma_2}e^{i\varphi} & \gamma_2 
\end{bmatrix}
\end{equation}
{The complex eigenenergies are derived as
\begin{equation}
\label{Edispersion}
\begin{split}
   \rm  E_{1,2}&= \rm \frac{E_1+E_2}{2}-i\frac{\hbar{(\gamma_1+\gamma_2)}}{2}\\&
   \rm  \pm \sqrt{[\frac{E_1-E_2}{2}-i\frac{\hbar(\gamma_1-\gamma_2)}{2}]^2+(\hbar\kappa-i\hbar e^{i\varphi}\sqrt{\gamma_1\gamma_2})^2}
\end{split}
\end{equation}}

 For the two identical monolayer metasurfaces, $E_1=E_2=E_0$, $\gamma_1=\gamma_2=\gamma_0$, the two complex eigenenergies of the stacked metasurfaces become
\begin{equation}{\label{eq:E}}
\begin{split}
&E_{1,2}=E_{\pm}=E_0\pm\hbar\kappa+ i\hbar\gamma_0 (1\pm e^{-i\varphi)}
\\&=E_0\pm\hbar(\kappa+\gamma_0 sin\varphi) 
\\&+i\hbar\gamma_0(1\pm cos \varphi).
\end{split}
\end{equation}
The corresponding resonant energy positions and linewidths of the two resonant modes can be determined by the real and imaginary parts of the two complex eigenenergies as Re$(E_{\pm})=E_0\pm\hbar(\kappa+\gamma_0 sin\varphi)$, Im$(E_{\pm})=\hbar\gamma_0 (1\pm cos\varphi)$, respectively. The $Q$ factors of the two resonant modes can be calculated as
\begin{equation}{\label{eq:Q}}
Q_{\pm}=\frac{1}{2}\frac{Re(E_{\pm})}{Im(E_{\pm})}=\frac{E_0\pm \hbar(\kappa+\gamma_0sin\varphi)}{2\hbar\gamma_0(1\pm cos\varphi)}.
\end{equation}
From the above equation,  resonant modes with infinite $Q$ are expected while $1\pm cos\varphi=0$. Correspondingly, $\varphi=n\pi$, $d=n\frac{\lambda}{2}$, which is in accordance with our previous discussion of reflectance and transmittance spectra based on the proposed theoretical model.

Considering further the mode splitting as shown in Fig.~\ref{coupling} (a), the energy difference of the two resonant modes can be calculated as $\Delta E=E_{+}-E_{-}=2\hbar(\kappa+\gamma_0sin\varphi)$. $\hbar\gamma_0$ represents the energy linewidth of a single metasurface.  In order to quantify the energy linewidth, the transmittance spectrum of a single  metasurface is shown in the inset of Fig.~\ref{coupling} (b). The transmittance spectrum can be estimated by the Fano equation, $T(E)=T_0+A_0\frac{[q+2(E-E_0)/\gamma_0]^2}{1+[2(E-E_0)/\gamma_0]^2}$ following Ref. \cite{fang1998determination},
where $E_0$ is the resonance energy and $E = hc/\lambda$, where $\lambda$ is the wavelength in free space, $T_0$ is the transmittance offset, $A_0$ is the continuum-discrete coupling constant, and $q$ is the Breit-Wigner-Fano parameter determining the asymmetry of the resonance profile. From the Fano fit, the resonance energy is $E_0=2.096$ eV ($\lambda$ = 591.6 nm) illustrated by the black line in the inset of Fig.~\ref{coupling} (b) and the energy linewidth of a single metasurface is determined to be $\hbar\gamma_0=0.012$ eV.

The resonance energy difference can be determined by the two minimum values of the transmittance spectra as shown in Fig.~\ref{coupling} (a), denoted as $\lambda_2$, $\lambda_2$ and correspondingly, $\Delta E=E_{+}-E_{-}=hc/\lambda_1-hc/\lambda_2=2\hbar(\kappa+\gamma_0sin\varphi)$ with $\varphi=k_0(s+t)$. The coupling strength is calculated for varying separations with the results shown in Fig.~\ref{coupling} (b). It is clear for separation $s<500$ nm, the coupling rate, $\kappa$ dominates the radiative damping rate, $\gamma_0$, defined as near-field regime. For the separation $s>500$ nm, the radiative damping rate, $\gamma_0$ becomes dominant and $\kappa$ becomes much weaker than $\gamma_0$ with increasing $s$, defined as the far-field regime.  This corroborates well with the observed degenerate resonant modes in the far-field, as shown in Fig.~\ref{coupling} (a), for $s$ = 550 nm, 600 nm and 650 nm, respectively. Furthermore, the near-field coupling is caused by the overlap of the evanescent resonant mode of each metasurface, which is in theory exponentially dependent on the separation between the two layers \cite{wu2022tailoring,luo2022wavy}. To further test this, the coupling strength versus $e^{-k_0 s}$ for $s<500$ nm is presented in Fig.~\ref{coupling} (c), the linear fit confirms the exponential dependence on the separation and further confirms that the near-field coupling dominates in this region. The weak coupling strength for separation $s>500$ nm indicates that each metasurface can be treated as a decoupled optical element in the far-field, which is an essential prerequisite of our model employing transfer matrix method or Eqs.~(\ref{eq:anar}) and ~(\ref{eq:anat}).

\subsection{Numerical validation and implication of the theoretical model}

To test our theoretical model, the simulated R, T curves of the two stacked metasurfaces are shown in Fig.~\ref{fig:fit} for the center-to-center distance, $d$, ranging from 1100 nm to 1350 nm with a step of 50 nm, together with the theoretical calculations according to Eqs.~(\ref{eq:anar}) and ~(\ref{eq:anat}). It is seen that the transfer matrix model quantitatively agrees with full-wave FDTD simulations. Fully numerical approaches are time-consuming, especially for a very sharp mode with a high $Q$, and they provide little insight. The transfer matrix approach developed in this work can therefore be used to theoretically predict spectral responses of the stacked metasurfaces geometry in the far-field coupling regime. 

A high-$Q$ resonant mode of the stacked metasurfaces with $d$ = 1180 nm is presented in Fig.~\ref{fig:highQ}, with the focus of unravelling the origin of the mode.  As can be seen in Fig.~\ref{fig:highQ} (a1), the simulated transmittance spectrum shows a sharp Fano-type resonant mode. The corresponding $Q$ factor is phenomenologically evaluated as $Q = E_0/(\hbar\gamma_0)$ = 2.09605 eV/(9.5$\times10^{-6}$) eV = 2.1$\times10^{5}$, which is approximately 2.2 $\times10^5$ times the Q-factor of the mode of a single metasurface, estimated from the transmittance spectra in the inset of Fig.~\ref{coupling} (b), with $Q = E_0/(\hbar\gamma_0)$ = 2.0961 eV/0.012 eV = 174.7. {To fully reveal the origin of the three order of magnitude increase of the high Q factor in the stacked geometry, the dependence of the calculated $Q$ factor (using Eq.~\ref{eq:Q}) on the center-to-center distance, $d$, is presented in Fig.~\ref{fig:highQ} (a2). It is clear the $Q$ factor becomes infinite where the distance meets the condition, $d=n\lambda/2$. The inspected distance for $d$ = 1180 nm is illustrated by the gray dash lines, which is consistent with the calculation from Fano fit.  Under such special conditions of metasurface separation, a Fabry-P\'{e}rot BIC is formed \cite{hsu2016bound,luo2022wavy}. 
In contrast to the previously reported examples of Fabry-P\'{e}rot BIC modes generated in photonic slabs or dielectric gratings, the Fabry-P\'{e}rot BIC modes generated by the stacked dielectric metasurfaces result from the coherent interplay of the Mie scattering resonant modes. }

Furthermore, the $Q$ factor enhancement accompanies a strong field concentration. To demonstrate this, the corresponding in-plane ($x-y$) field enhancement in the center of M1 in both cases, which is the amplitude ratio, $|E/E_0|$ are shown in Figs.~\ref{fig:highQ} (a3) and (b1) respectively. The inspected wavelength is 2.09606 eV ($\lambda$= 591.6 nm) for both cases, which is the transmittance maximum of the stacking metasurfaces with the maximum field concentration. The electric field vectors for both cases are also shown. It is clear that the electric field distribution and the electric field vectors in M1 for the isolated case and the stacked case display similar maps, which corroborates well with the theoretical assumption of negligible near-field coupling between the two metasurfaces at the inspected separation. However, the amplitude ratio of the electric field amplitude in $x-y$ plane for the stacked case shows an approximate 20-fold enhancement at the inspected wavelength compared to that of the M1 alone, referred to as the supercavity resonance wavelength. Due to the negligible near-field coupling of the stacked metasurfaces over such a large separation, the higher field concentration results only from radiative damping and propagation phase, the standing wave is generated and the energy is stored in the Fabry-P\'{e}rot cavity. The standing wave can be further confirmed by the out-of-plane field map ($x-z$) in Fig.~\ref{fig:highQ} (a3), which records the electric field amplitude ratio along $z$ direction, illustrating an approximately 35-fold electric field amplitude enhancement within the Fabry-P\'{e}rot cavity in free-space between two metasurfaces compared to the incident light wave. 

It is further interesting to note that as in Fig.~\ref{fig:highQ} (a1), the transmittance of the isolated metasurace at the supercavity's resonance wavelength is 17$\%$, corresponding to a reflectance of 83$\%$ for the lossless metasurface. The non-zero and non-unity transmittance at the supercavity's resonance wavelength are the prerequisite for the Fabry-P\'{e}rot mode. 17$\%$ of the incident wave propagates into the cavity and generates the standing wave due to 83$\%$ of the reflectance at both metasurfaces. The standing waves coherently overlap with the scattering Mie-type multipolar modes generated in M1 and M2, generating the high-$Q$ supercavity mode.

\section{Summary}\label{sec4}

To summarize, we theoretically explore the electromagnetic responses of two stacked identical metasurfaces. Through analytical calculations based on multipole decomposition, transfer matrix method, temporal coupled mode theory and FDTD simulations, it is demonstrated that the electromagnetic responses of the stacked metasurfaces in the far-field can be well reproduced from the Mie-resonant modes of the monolayer metasurface. The theoretical results deepen our understanding of light manipulation driven by the coherent interplay of Mie-type radiative scattering modes and the Fabry-P\'{e}rot mode. Based on this, we have presented a planar metasurface cavity. High-$Q$ supercavity modes with the in-plane and out-of-plane field confinement can be created via the Fabry-P\'{e}rot bound state in the continuum. We envision that, this approach could be extended to other geometries, such as ultra-thin  metasurfaces composed of arbitrarily shaped dielectric nanoparticles and for structures embedded in a dielectric medium. From the perspective of applications, the designed supercavity mode could be easily accessed for the integration of quantum emitters inside the cavity, which may enrich the exploration of lasing devices and is potentially useful in nonlinear physics,  quantum correlations and optical trapping.

\section{Appendix}

\begin{figure}[h]
\includegraphics[width=0.8\linewidth]{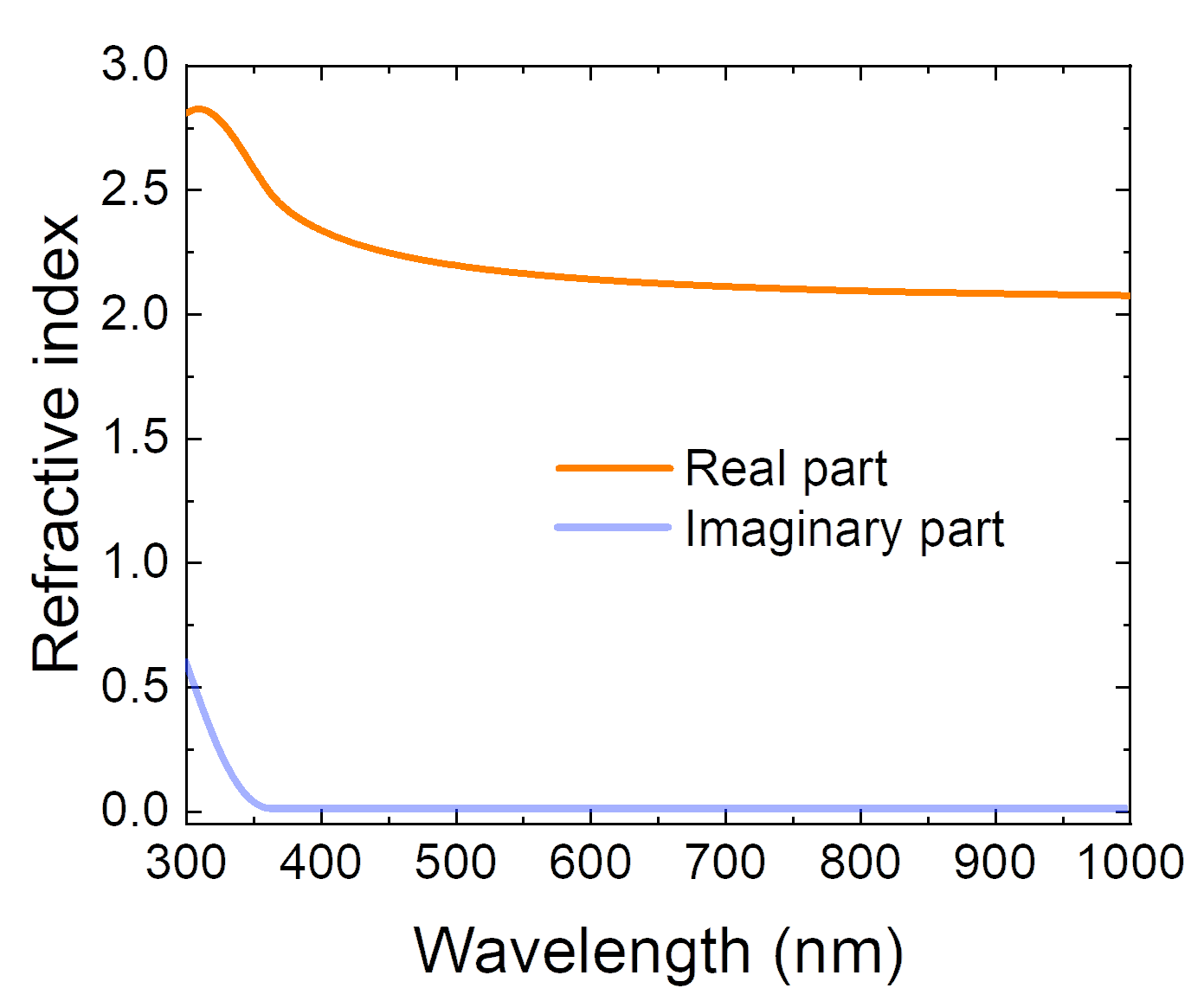}
\caption{\label{n} Real and imaginary part of the refractive index of TiO$_2$ following Ref. \cite{Sarkar2019}.}
\end{figure}

\begin{acknowledgments}
This work was supported by the National Natural Science Foundation of China (Grant No. 11975072) and Science Foundation Ireland (SFI) (Grant No. SFI-21/FFP-P/10187).
\end{acknowledgments}

\bibliography{apssamp.bib}

\nocite{*}

\end{document}